# Title

## Effects of ligand binding on the energy landscape of acyl-CoA-binding protein

Punam Sonar[1], Luca Bellucci[2], Alessandro Mossa[3,4]*, Pétur O. Heidarsson[5]*, Birthe B. Kragelund[6]*, and Ciro Cecconi[7,8]*.

[1]*Physik-Department E22, Technische Universität München, James-Franck-Str. 1 85748 Garching Germany;*

[2]*CNR-NANO, CNR Institute Nanoscience, NEST Scuola Normale Superiore, Piazza San Silvestro 12, 56127 Pisa, Italy;*

[3]*INFN Firenze, Via Giovanni Sansone 1, 50019 Sesto Fiorentino, Italy;*

[4]*Istituto Statale di Istruzione Superiore "Leonardo da Vinci", Via del Terzolle 91, 50127 Firenze, Italy;*

[5]*Department of Biochemistry, Science Institute, University of Iceland, Dunhagi 3, 107 Reykjavík, Iceland;*

[6]*Structural Biology and NMR Laboratory, Department of Biology, University of Copenhagen, Ole Maaløes Vej 5, 2200 Copenhagen N, Denmark;*

[7]*Department of Physics, Informatics and Mathematics, University of Modena and Reggio Emilia, 41125 Modena, Italy;*

[8]*Center S3, CNR Institute Nanoscience, Via Campi 213/A, 41125 Modena, Italy;*

*Correspondence: pheidarsson@hi.is (P.O.H.), bbk@bio.ku.dk (B.B.K.), alessandro.mossa@gmail.com (A.M.), ciro.cecconi@gmail.com (C.C.)

**Statement of Significance**

Many proteins interact with one or more ligands to perform their biological functions. Yet, the effects of ligand binding on the conformational energy landscapes and thus on the mechanical properties of proteins are still largely unknown. Here we use optical tweezers and molecular dynamics simulations to elucidate the effects of octanoyl-CoA on the mechanical resistance and compliance of acyl-CoA binding protein (ACBP). Based on our results we provide an in-depth discussion on the functional significance of the conformational energy landscape of holo-ACBP. The results of this work reveal new mechanisms of interaction between a protein and its ligand, and suggest novel strategies for modulating the mechanical properties of proteins.

**ABSTRACT**

Binding of ligands is often crucial for function yet the effects of ligand binding on the mechanical stability and energy landscape of proteins are incompletely understood. Here we use a combination of single-molecule optical tweezers and MD simulations to investigate the effect of ligand binding on the energy landscape of acyl-coenzyme A (CoA) binding protein (ACBP). ACBP is a topologically simple and highly conserved four-α-helix bundle protein that acts as an intracellular transporter and buffer for fatty-acyl CoA and is active in membrane assembly. We have previously described the behavior of ACBP under tension, revealing a highly extended transition state (*TS*) located almost halfway between the unfolded and native states. Here, we performed force-ramp and force-jump experiments, in combination with advanced statistical analysis, to show that octanoyl-CoA binding increases the activation free energy for the unfolding reaction of ACBP without affecting the position of the transition state along the reaction coordinate. It follows that ligand binding enhances the mechanical resistance and thermodynamic stability of the protein, without changing its mechanical compliance. Steered molecular dynamics simulations allowed us to rationalize the results in terms of key interactions that octanoyl-CoA establishes with the four α-helices of ACBP and showed that the unfolding pathway is marginally affected by the ligand. The results show that ligand-induced mechanical stabilization effects can be complex and may prove useful for the rational design of stabilizing ligands.

**INTRODUCTION**

Molecular forces are involved in almost all stages of a cell's life cycle, making the mechanical properties of proteins biologically relevant and key to understand force regulated cellular processes. Many proteins have a direct force bearing function, such as proteins involved in maintaining the structural integrity of cells (1-4), molecular motors that convert chemical energy into kinetic energy to generate motion (5, 6) and mechano-sensors that respond to cellular mechanical stress (7, 8). Understanding the effects of force impacts the design of nanomaterials and the rational design of ligands that can modulate the mechanical properties of proteins, which is relevant for drug development. Especially for the latter, studying how natural ligands affect the mechanical properties of their protein partners is critical.

Optical tweezers have recently emerged as a powerful tool to monitor and even modulate, in a highly controlled fashion, the conformational state of a protein molecule as a function of force and end-to-end distance (9-11). Through the use of DNA molecular handles, the molecule is tethered between two beads and by changing the distance between the beads, the conformational fate of the molecule can be controlled by the applied tension (12, 13). Force-extension traces, as well as (un)folding forces and rates are directly accessible in these experiments. The information they contain can be used to reconstruct the salient features (or, in some cases, even the full profile) of the (un)folding free energy landscape (14).

Relatively few proteins have been studied by mechanical manipulation thus far (11, 15-25) and even fewer have been studied for the effects of ligand binding (26-29). A ligand may affect the mechanical properties of a protein through several potential mechanisms involving specific ligand-protein interactions and minor or extensive conformational rearrangements which will modulate the energy landscape. Yet, no correlation between binding affinity, induced structural changes and changes in mechanical properties of the protein has so far been revealed. Rather, a broad spectrum of behaviors has been described, revealing the complexity of ligand-protein interactions. In some cases ligand binding has a moderate (30, 31) or even no apparent (32, 33) effect on the mechanical properties of a protein; in other cases it significantly reduces the unfolding rate constant and thus increases the average protein unfolding force (34, 35). Similarly, some ligands lead to additional intermediate states along the unfolding pathways of a protein (36, 37), while in other cases they stabilize preexisting intermediate conformations (30, 37). Thus, a better understanding of the principal molecular mechanisms mediating the effect of a ligand on the energy landscape of a protein is clearly needed.

In this work, we characterize the effect of ligand binding on the folding energy landscape of acyl-coenzyme A (CoA) binding protein (ACBP). ACBP is an 86-residue four-helical bundle protein of the ACBP-family, and it is active in cellular acyl-CoA transport and pool formation as well as required for fatty acid chain elongation and sphingolipid synthesis (38). ACBP binds acyl-CoAs of different lengths, with preference for long chain acyl-CoAs (39, 40). Due to its small size and relatively simple topology, ACBP has over the past 20 years been used extensively as a model system to study protein folding using various biophysical techniques (41-44). Previous mechanical unfolding experiments have shown that ACBP has an unfolding transition state that is unusually extended and independent of the force vector (pulling direction) that is applied (45). The wealth of information previously obtained by optical tweezers on the mechanical (un)folding of ACBP and by bulk studies on its ligand binding properties gives us a powerful platform to decode the effects of ligand binding on the energy landscape of this protein. By applying force across the entire polypeptide chain and using statistical analysis on the data, we found that single ACBP molecules undergo a distinct ligand-induced mechanical stabilization. We rationalize the results using steered-molecular dynamics simulations and provide evidence that the unfolding mechanism of the protein is only slightly affected by the presence of ligand despite an increased height of the transition state barrier.

**RESULTS and DISCUSSION**
**Single-molecule force spectroscopy of ACBP in the presence of its ligand**
To secure a solvent accessible ligand for the single-molecule studies we used octanoyl-CoA, which binds to ACBP with low-micromolar affinity with a $K_D$ of 0.33μM (40). Single ACBP molecules were mechanically manipulated with optical tweezers as depicted in Figure 1. Force was applied to the N and C termini of the protein through DNA molecular handles covalently attached to terminal cysteine residues engineered at position 1 and 86, effectively probing the force response of the entire polypeptide chain. The unfolding and refolding processes of ACBP in the presence of octanoyl-CoA were investigated through both force-ramp and force-jump experiments, as previously done for apo-ACBP (45). Force-ramp experiments were performed by moving the pipette relative to the optical trap at constant speed to stretch and relax the molecule at almost constant loading/relaxation rate, which generated Force-Extension Curves (FECs) as in Figure 1B (46-48). We observed sudden changes in the extension of the molecule (transitions) during both stretching (at ~ 12 pN) and relaxation (at ~ 4 pN ) corresponding respectively, to the full unfolding and refolding of ACBP in a two-state manner, as indicated by the changes in contour length ($\Delta L_c$) associated with these transitions (Figure 1B). Once tethered between the two polystyrene beads, individual ACBP molecules were stretched and relaxed multiple times to generate hundreds of FECs, which were then used to obtain distributions of unfolding and refolding forces. Under saturating conditions of octanoyl-CoA (40) (Materials and Methods), ACBP becomes mechanically more resistant than the apo form, unfolding at ~ 12 pN, but refolds at the same forces as in the absence of ligand (45) (Figure 1C and 1D), suggesting ligand binding to occur subsequent to folding, as seen previously in bulk with the ligand palmitoyl-CoA (PCoA) (49). To extract information on the energy landscape of the protein, unfolding and refolding force distributions were analysed using the Bell model that postulates that the unfolding ($k_u$) and refolding rate ($k_f$) constants depend exponentially on force (50):

$$k_u(F) = k_m k_u^0 \exp(F\, x^{\ddagger}_u / k_B T) \tag{1}$$
$$k_f(F) = k_m k_f^0 \exp(-F\, x^{\ddagger}_f / k_B T) \tag{2}$$

where $k_m$ is a constant that reflects any contribution from the "machine" (such as handles and beads) to the measured rates, $k_u^0$ and $k_f^0$ are the rate constants of unfolding and refolding of the molecule at zero force, $F$ is the applied force, $x^{\ddagger}_u$ and $x^{\ddagger}_f$ are the distances from the TS to the folded (N) and unfolded (U) states along the reaction coordinate and $k_B T$ is the product of the Boltzmann constant and the absolute temperature. For holo-ACBP, the analyses yielded a distance from TS to N ($x^{\ddagger}_u$) of 5.8 ± 0.4 nm, a distance from TS to U ($x^{\ddagger}_f$) of 6.8 ± 0.4 nm, $k_m k_u^0 = 6.0(\pm 4)\cdot 10^{-7}$ s$^{-1}$ and $k_m k_f^0 = 1(\pm 0.4)\cdot 10^3$ s$^{-1}$, (Figure 2, Table 1). As an independent and complementary analysis, unfolding and refolding events of holo-ACBP were also analyzed at constant force by performing force-jump experiments. In these experiments, the force is rapidly increased (jumped) or decreased (dropped) to preset values and then held constant (clamped) until an unfolding or refolding event is observed (45, 51). The dwell-time distributions of the folded and unfolded states of the protein obtained at different forces are then used to correlate ln$k$ to force (Figure 2D), which can be analyzed with the Bell model to extract information on TS and rate coefficients. This analysis yielded: $x^{\ddagger}_u = 6.0 \pm 0.73$ nm, $x^{\ddagger}_f = 7.9 \pm 0.73$ nm, $k_m k_u^0 = 2.9(\pm 5.3)\cdot 10^{-7}$ s$^{-1}$ and $k_m k_f^0 = 5.4(\pm 4.9)\cdot 10^3$ s$^{-1}$ (Table 1). These values are in agreement with those obtained through force-ramp experiments validating the accuracy of the two independent studies. The $x^{\ddagger}_u$ value tells us the amount of elastic deformation a protein can undergo along the pulling axis before crossing the unfolding transition state barrier, and thus tells us how compliant a protein is. In order to compare the mechanical compliance of different proteins, the "mechanical" Tanford $\beta$-value ($m\beta_{Tu}$) was introduced that provides a normalized distance to the transition state as the ratio between $x^{\ddagger}_u$ and the N to U distance ($x^{\ddagger}_u/\Delta x$) (52). The average $m\beta_{Tu}$ of 0.45 ± 0.7 that emerges from our studies is in agreement with that measured for the apo form (0.45 ± 0.06) (45), indicating no detectable effect of ligand binding on the position of the *TS*.

Rate constants estimated through optical tweezers manipulation studies contain contributions from experimental parameters ($k_m$) and thus it is difficult to compare them to intrinsic rate constants of the protein. However, since the experimental conditions (optical tweezers setup, beads and molecular handles) used here for holo-ACBP are identical to those used for apo-ACBP (45), the transition rates can be directly compared. This comparison reveals that while the refolding rate constants measured in the presence of octanoyl-CoA (Table 1) are quite similar to those measured for the apo-form ($k_m k^0_f$ = 3.8(± 0.8)·$10^3$ $s^{-1}$ and 4.25(±0.7)·$10^3$ $s^{-1}$ in force-ramp and force-jump experiments, respectively[*]), the unfolding rate constants (Table 1) are two orders of magnitude lower than in the absence of ligand ($k_m k^0_u$ = 2.8 (±1.0)·$10^{-5}$ $s^{-1}$ and 3.1 (±2.5)·$10^{-5}$ $s^{-1}$ in force-ramp and force-jump experiments, respectively[*]) indicating a higher unfolding activation barrier for holo-ACBP. Consistently, holo-ACBP is mechanically more resistant than the apo form but refolds at the same forces (Figure 1C and 1D). The observation that the ligand has no effect on the refolding forces and rates of ACBP, suggests that it binds only after formation of the native state and has negligible affinity for the unfolded state. This is in agreement with bulk experiments showing negligible effect of the ACBP ligand S-hexadecyl-CoA on the rate of formation of the hydrophobic core of the protein (53) and with bulk amide hydrogen-to-deuterium exchange experiments of apo- and holo-ACBP, identifying a generally stabilized core of ACBP from ligand binding (54).

**Free energy estimates from non-equilibrium experiments**
The FECs collected in the force-ramp experiments can be used to estimate the free energy difference *ΔG₀* between the fully unfolded and native states by means of the celebrated Crooks' fluctuation theorem (55). Though this procedure is well established, there are technical issues related to the data analysis that have not been satisfyingly addressed in the past. One issue is the use of the end-to-end molecular extension in lieu of the proper control parameter, which is the distance between the center of the trap and the tip of the pipette, a quantity not easily accessible to our instrument. The low bandwidth of our data acquisition system and the use of a bi-directional method (as opposed to the Jarzynski equality (56), which uses only the "forward" data) make it possible to replace the total distance with the bead-to-bead extension without spoiling the applicability of the fluctuation theorem (57). The second issue is how to correctly remove the effect of the DNA handles in order to estimate the free energy of the protein alone. In this paper, we introduce a novel method based on a previous approach (58), which entails the reconstruction of the Thermodynamic Force-Extension Curve (TFEC), i.e. the FEC that we would observe if we were able to perform the experiment under reversible conditions and take an average over infinite realizations (see Figure 3, and Materials and Methods for details about the data analysis procedure). Strictly speaking, the TFEC, defined as the mean force as a function of the mean extension in the statistical ensemble where the total distance trap-pipette is held fixed, does not present such sharp angles at the extremities of the rip as those showed in Fig. 3. However, the TFECs computed from realistic models of single-molecule experiments (58) are similar enough to our reconstructed curve as to warrant a minor abuse of nomenclature.

---
[*] unpublished data from the experiments reported in (45).

From the reconstructed TFEC (more precisely, from the region of the rip), we can estimate several useful quantities related to the unfolding transition: the coexistence force $f_c$, the extension $\Delta x_c$ released upon unfolding, and the zero-force free energy of unfolding $\Delta G_0$ (see Materials and Methods for details about how $\Delta G_0$ is estimated). Our results for five datasets are summarized in Table 2. By performing averages weighted over the uncertainties, we obtain the following estimates: $\Delta x_c = (11.2 \pm 0.4)$ nm, $f_c = (7.4 \pm 0.3)$ pN, and $\Delta G_0 = (14.1 \pm 1.2) k_B T = (8.3 \pm 0.7)$ kcal/mol. This is the free energy difference between the fully denatured apo ACBP and the native holo ACBP states. The same force-ramp experiments have been carried out under the same experimental conditions except for the absence of the ligand: the results are summarized in Table 3. A weighted average over four datasets yields $\Delta x_c = (8.1 \pm 0.2)$ nm, $f_c = (5.8 \pm 0.2)$ pN, and $\Delta G_0 = (6.9 \pm 0.4) k_B T = (4.0 \pm 0.2)$ kcal/mol. This is the free energy difference between the fully denatured apo ACBP and the native apo ACBP states. The stabilizing effect of the ligand is clear and can be quantified as $\Delta\Delta G_0 = (7.2 \pm 1.3) k_B T$, that is the free energy difference between the native apo ACBP and the native holo ACBP states. To assess the structural determinants of the stabilizing effect, we turned to molecular simulations.

**Ligand binding does not change the unfolding mechanism**

To investigate the effect of the ligand octanoyl-CoA on the mechanical denaturation of ACBP at atomic level, we performed all-atom standard MD and steered molecular dynamics (SMD) simulations in explicit solvent. The holo form of ACBP was modeled starting from the NMR structure of the protein bound to palmitoyl-CoA (PDB code 1NVL); the octanoyl-CoA was reconstructed cutting the atoms in excess from the palmitoyl chain (Figure S3). Apo-ACBP was modelled from the NMR structures 1NVI. The apo- and holo-ACBP structures were first relaxed through a 50 ns unrestrained MD simulation (Materials and Methods, Figure S4). Then the energetics of ligand binding was characterized through a pairwise decomposition of residue-ligand potential interaction energies. The averaged computed potential energies range from 5 to -180 kcal/mol and are shown by color scale in figure 4B. Out of the four ACBP helices (H1-H4), and similar to the structure of the complex between ACBP and PCoA (49), octanoyl-CoA binds mostly to residues of H2 and H3, and partially to residues of H1 and H4 (Figure 4 A-B). In fact, H1 interacts with the ligand with its C-terminal through Val12 and Lys13. Val12 interacts with the acetyl group and the first part of the octanoyl-CoA chain, whereas Lys13 establishes an electrostatic interaction with the negatively charged pyrophosphate group. Lys18 of the H1-H2 loop also interacts with octanoyl-CoA, but this electrostatic interaction is partly shielded by the solvent and likely plays a minor role into the protein stabilization. H2 interacts with the ligand through both hydrophobic and electrostatic interactions. Met24 and Ile27 establish hydrophobic interactions with the first part of the octanoyl chain, while Tyr28 forms a stable hydrogen bond with the phosphate of the 3'-phosphoadenosine group. Tyr31 binds to the adenosine group with $\pi$-stacking interactions, whereas Lys32 forms an ionic interaction with the phosphate of the 3'-phosphoadenosine group. H3 interacts with the phosphate of the 3'-phosphoadenosine group mainly with Lys54. Lys50 shows a significant electrostatic interaction with the ligand but being solvent-exposed its contribution to the binding stabilization is unlikely to be relevant. Remarkably, H4 interacts with the ligand only with Tyr73 by means of two hydrogen bonds established between the hydroxyl group of the tyrosine and the adenosine group of octanoyl-CoA.

The atomistic details of the mechanical denaturation of the apo and holo forms of ACBP were investigated through SMD simulations (Materials and Methods). The two termini of the protein were attached to an ideal spring with an elastic constant of 0.25 kcal mol$^{-1}$ Å$^{-2}$, and they were pulled apart at a speed of 10 Å/ns. Structural changes of the protein during the simulation were monitored as described before (59). The sequences of unfolding events of the apo- and holo-ACBP are remarkably similar (Figure 4C and video available online at ....). In both cases, ACBP begins to unfold from H1 that loses first its tertiary contacts with the rest of the protein and then its intrahelical contacts, unraveling in an elongated polypeptide chain. Then H4 starts losing its tertiary contacts. Finally, H2 and H3, which are subject directly to force only after H1 and H4 undocking, are the last to yield. It is interesting to observe that this is the same picture of unfolding under tension that was established for the apo form of ACBP using ratcheted molecular dynamics (45): such agreement between quite different simulation techniques is an important validation and an incentive to take seriously this description as a basis for further interpretation of our experiments.

As shown in Figure 4C, the ligand seems not to affect the unfolding trajectory of ACBP, although it increases its unfolding activation barrier, as described above. This peculiar effect of octanoyl-CoA on the protein denaturation mechanism could have its origin in the way it interacts with the different parts of ACBP and in the structure of the unfolding transition state. As shown in Figure 4B, octanoyl-CoA interacts mostly with H2 and H3, which have been suggested to constitute the structured part of the unfolding transition state of ACBP (45, 60), and only marginally with the rest of the protein. We speculate that overall octanoyl-CoA binding stabilizes the native state of ACBP more than its unfolding transition state, thus increasing the unfolding activation barrier and the mechanical resistance of the protein. However, the few and weak interactions that the ligand forms with H1 and H4 affect only marginally the molecular events preceding the crossing of the *TS* barrier in apo-ACBP, i.e. the denaturation of H1 and the detachment of H4 from H2 and H3, thereby leaving the unfolding trajectory substantially unchanged and $x^{\ddagger}_u$ unaffected (Table 1). This is consistent with the observation from hydrogen-deuterium exchange measurements comparing apo- and holo-ACBP, which show that although stabilized by ligand binding, H1 remains the least stable helix of ACBP (54).

**CONCLUSIONS**

Some proteins can couple binding with folding, as is prominent in the interactions of many intrinsically disordered proteins (61-63), but this does not seem to be the case for the unfolded state of ACBP. The presence of octanoyl-CoA has no effect on the refolding rates and forces, suggesting binding of the ligand only after full refolding of the protein, as also suggested by bulk experiments using a ligand with a longer acyl-chain (53). Nonetheless, ACBP becomes mechanically more resistant and thermodynamically more stable in the presence of octanoyl-CoA, unfolding at higher forces. Yet, the unfolding pathway of the protein is only marginally affected by the ligand which enhances the height of the transition state barrier without changing its position along the reaction coordinate. In fact, in our SMD simulations the sequences of unfolding events of the apo- and holo-ACBP are remarkably similar. In both cases, ACBP unfolding starts with the unraveling of H1 that soon becomes an elongated polypeptide chain, losing both its tertiary and intrahelical contacts. Then H4 detaches from the rest of the protein and unfolds while H2 and H3 lose their structures only later, when the protein has been stretched considerably. As the position of *TS* does not change upon ligand binding, it follows that the large pliability of the protein is not affected. This allows ACBP to go through significant deformation even when transporting cargo. In that way, ACBP is quite a robust molecule that can be subjected to significant deforming forces without losing its ligand. This might be important partly because of the need to accommodate ligand with different acyl chain length, for delivery of acyl-CoA to targets including vesicles, for translocation or for interactions with other protein targets in ternary complexes. Our previous work (45) prompted us to speculate that this compliance might even be an evolutionary-tuned feature that allows proteins that need to be translocated or unfolded to consume less cellular energy as increased compliance makes the unfolding rates more sensitive to force. Translocation of polypeptides through ClpX proteolytic *E. coli* machinery proceeds through mechanical unfolding and the translocation kinetics were sensitive to pulling direction and mechanical stability (64) whereas ACBPs pliability is rather insensitive to pulling direction (45). It was recently demonstrated that the mechanical stability of transcription factors regulates their translocation rate into the nucleus (65). Interestingly, ACBP has been found to interact with hepatocyte nuclear factor-4 alpha, which regulates the transcription of genes involved in both lipid and glucose metabolism, in the nucleus of intact cells (66). Transport across the central pore of the nuclear pore complex is finely regulated by intrinsically disordered nucleoporins and it is believed that an extended polypeptide more efficiently overcomes an entropic barrier conferred by the disordered nucleoporins than a stiffer native structure does (67). Furthermore, ACBP has in human cells been shown to be secreted into the extracellular space upon starvation (68), possibly as a response to the increased need for fatty acid synthesis in other organs, which may require many barrier-crossing steps where partial unfolding may be required.

Our results here show that the high pliability, and thus sensitivity to force, is also extended to the ACBP/acyl-CoA complex. Partial unfolding of lipid-bound ACBP may be sufficient and even beneficial for translocation over a membrane without full unbinding of the ligand. Given that ACBP binds acyl-CoA esters with a broad range of acyl-chain lengths (C8–C24) (39) and the acyl-chain forms part of the binding site for the CoA headgroup, ACBP seems to have evolved to have a built-in pliability that may be important for ligand binding and delivery, across cellular compartments and between cells. Overall, our results shed light on the effects of ligand binding on the energy landscape of ACBP and open the prospect for the rational design of small molecule ligands that increase the mechanical stability of proteins without altering their native structure or their unfolding mechanism.

## MATERIALS AND METHODS
### Protein expression, purification, and sample preparation
The double cysteine variant ACBP$_{C1C86}$ was expressed in *Escherichia coli* BL21(DE3)-pLysS cells transformed with a pET3a expression vector containing the mutated bovine ACBP gene (69). Purification was performed as previously described (70). The DNA-protein coupling reaction to generate DNA-protein chimeras for use in optical tweezers experiments was performed exactly as described (48).

### Optical tweezers experiments

Experiments were performed using a custom-built optical tweezers instrument with a dual-beam laser trap (45). DNA-protein constructs were tethered between two polystyrene beads. A 3.10 μm antidigoxigenin-coated bead (Spherotec) held in the optical trap and a 2.18 μm streptavidin-coated bead (Spherotec) held at the end of a micropipette by suction. Force was applied on the protein by moving the micropipette relative to the optical trap using a piezoelectric flexure stage (MAX311/M, Thorlabs, Newton, NJ). Measurements were conducted at ambient temperatures in 10 mM Tris-HCl buffer, 250 mM NaCl, 10 mM $CaCl_2$, 0.04% $NaN_3$, in the absence or presence of 44 μM octanoyl-CoA (Avanti Polar Lipids, Inc. 870708P 2 Mg), pH 7.0. The force applied on the protein was determined by measuring the change in momentum flux of the light beams leaving the trap, while changes in the extension of the molecule were determined by video microscopy (71). During force-ramp experiments, the pipette was moved at constant speed (nm $s^{-1}$). Under these experimental conditions, above ~3 pN, force changed approximately linearly with time and thus the loading rate (pN $s^{-1}$) was approximately constant. Data were collected only on molecules that showed the characteristic overstretching transition at 67 pN of force (72). In force-ramp experiments data were recorded at a rate of 40 Hz. In force-jump experiments, the force applied on the molecule was jumped between two set-point values and kept constant through a force-feedback mechanism. The dead-time (time of jump) was measured to be 60 ms. In force-jump experiments, the data were acquired at a rate of 100 Hz.

**Molecular dynamics simulations**

Molecular dynamics simulations were performed with NAMD (v2.10)(73). The system preparation was done with VMD (v1.9.2) (74). The CHARMM27 force field (75) was used for the protein, octanoyl-CoA and the counterions, whereas the TIP3P (76) force field was used for water. Apo-ACBP was modeled starting from the first model of the refined NMR structural ensemble deposited in the protein data bank with the PDB code 1NTI. The holo form of ACBP was modeled starting from the first model of the NMR structural ensemble of the protein bound to palmitoyl-CoA (PDB code 1NVL). Remarkably, the presence of the ligand does not affect the main structure of ACPB being the backbone RMSD value between the NMR structures of the apo and holo forms lower than 2 Å, Figure S4. Octanoyl chain was modeled by cutting the last terminal atoms of the palmitoyl chain (figure S3). The apo and holo forms of ACBP were surrounded by a periodic box of water molecules. The water layer separating the protein was at least 10 Å thick. The neutrality of the apo- and holo-ACBP systems was guaranteed by adding 2 and 5 $Na^+$ ions, respectively. The lengths of all bonds involving hydrogen atoms were constrained using the SHAKE algorithm (77). The r-RESPA multiple time step method was employed with 2 fs for bonded, 2 fs for the short-range part of the non-bonded, and 4 fs for the long-range part of the electrostatic forces (78). MD simulations for the apo and holo systems were conducted using periodic boundary conditions (PBC) and the long-range part of the electrostatic was treated with the Particle-Mesh-Ewald (PME) method (79). The distance cut off for non-bonded interactions was set to 10 Å, and the switching function was applied to smooth interactions between 9 and 10 Å. MD simulations were conducted in the NPT ensemble. The temperature was set to 310 K and the Langevin thermostat was employed for temperature regulation whereas pressure was set to 1 atm and regulated via isotropic Langevin piston manostat as implemented in NAMD software package (73). A first minimization (2000 steps of conjugate gradient) to eliminate bad atomic contacts was followed by 500 ps of a position-restrained MD simulation. The obtained structures were then minimized without restraints and the final conformations were then subjected to 20 ns MD equilibration. Finally, 50 ns of production simulations were performed for both systems, saving conformations every 1 ps. Root mean square deviation (RMSD) based clustering over $C_\alpha$ atoms was performed with a Wordom software package (80) using a threshold of 2 Å. Cluster analysis of the trajectories followed by the apo and holo forms during the 50 ns production simulation revealed that the protein structures of the most populated clusters do not significantly differ, being the RMSD values between them not larger than 1.6 Å. Potential interaction energies between amino acid residues of ACBP and the octanoyl-CoA were evaluated post processing the last 50 ns of trajectory by using NAMD and the same force field parameter used for the MD simulation. Potential energy consisted of the sum of Lennard-Jones and the real part of the electrostatic interaction.

Steered molecular dynamics (SMD) simulations were performed starting from a series of selected conformations of the systems generated during the production phase. The apo- and holo-ACBP systems were both modeled and simulated using the same parameters used for the unrestrained MD, unless explicitly stated. For each selected conformation, the protein was centered into the box and rotated to place the $C_\alpha$ atoms of the first and the last residue along the *x*-axis. SMD simulations were conducted in the NVT ensemble without applying PBC and PME. The distance cut off for non-bonded interactions was set to 12 Å, whereas the switching function was applied between 10.5 and 12 Å. A preliminary protocol of minimization and 20 ps of MD simulations were then performed maintaining the protein fixed to the original position. This protocol was repeated six times, assuring that the freely moving water molecules gradually relaxed around the protein to form a "bubble" as previously described (81). At the end of the protocol, a further minimization and 20 ps of MD simulations were performed without any restraints in the atomic positions to relax the protein. The final conformations were then used as starting point for the SMD simulations. SMD simulations were performed by restraining the $C_\alpha$ atom of residue 1 to its initial coordinates and applying a force to a dummy atom attached to the C$_\alpha$ atom of residue 86 via a virtual spring with an elastic constant of 0.25 kcal mol$^{-1}$ Å$^{-2}$. The pulling velocity was set to 10 Å/ns along the *x* direction. A total of 32 SMD simulations were performed for both systems starting from structures saved during the last 32 ns of the 50 ns production simulation.

The video available online at …. shows SMD simulations of the mechanical unfolding of the apo and holo forms of ACBP, top and bottom panels respectively. In both cases the protein is stretched at 10 Å/ns to a final extension of 13 nm.

**Position of *TS* and rate constants from force spectroscopy experiments**

When a protein unfolding in a two-state manner is subjected to a force increasing linearly with time, the fraction (*N*) of folded molecules at the force *F* and loading rate *r* (pN s$^{-1}$) is given by (50, 82):

$$N(F,r) = exp[(k_m k_u^0 k_B T / r x^\ddagger_u) \cdot (exp(x^\ddagger_u F / k_B T) - 1)] \qquad (3)$$

In the high force limit (*F* > 3 pN), the exponential term is large compared to 1 and this equation can be simplified and linearized as:

$$ln[r\ ln[1/N(F,r)]] = ln(k_m k_u^0 k_B T / x^\ddagger_u) + (x^\ddagger_u F / k_B T) \qquad (4)$$

where *N(F,r)* can be calculated by integrating the unfolding force distributions over the corresponding force range and equation 4 can be used to fit *ln[r ln[1/N(F,r)]] vs F* graphs (Figure 2C) to estimate $k_m k_u^0$ and $x^\ddagger_u$ as best fit values.

With similar considerations we obtain the equation:

$$ln[-r\ ln[1/U(F,r)]] = ln(k_m k_f^0 k_B T / x^\ddagger_f) - (x^\ddagger_f F / k_B T) \qquad (5)$$

that can be used to fit *ln[-r ln[1/U(F,r)]] vs F* graphs (Figure 2C) to estimate $k_m k_f^0$ and $x^\ddagger_f$ as best fit values. The fraction (*U*) of unfolded molecules at the force *F* and loading rate *r* can be calculated by integrating the refolding force distributions over the corresponding force range.

**Thermodynamic force-extension curve (TFEC) reconstruction**

The reconstruction of equilibrium free energy differences between different molecular states by using nonequilibrium force-ramp experiments is customarily achieved by applying Crooks' fluctuation theorem

$$P_f(W)/P_r(-W) = \exp[(W-W_{rev})/k_BT] \qquad (6)$$

In the formula, $P_f(W)$ and $P_r(W)$ are the probability distributions of the work $W$ made on the system during the "forward" process (i.e., the stretching) and the "reverse" process (i.e., the relaxation), respectively. The reversible work $W_{rev}$ is the free energy variation of the entire system (tethered protein plus DNA linkers). Although the use of Eq. 3 is by now widespread in the optical tweezers community, it should be noted that the proper definition of work in a force-ramp experiment would require the measurement of the actual control parameter, which is in our case the total distance between the center of the trap and the tip of the pipette, a quantity not readily accessible in our setup. This question has been exhaustively treated elsewhere (57), here suffice to say that in the case of our experiment, the error due to the substitution of the control parameter with the molecular extension (the distance between the beads) is small compared to other sources of uncertainty.

We therefore compute the work associated with each trace as the area below the FEC, obtaining unfolding and refolding work distributions such as those showed in Figure S1. According to Eq. 3, the two distributions meet at the value of the reversible work $W_{rev}$; in practice, it is more efficient to apply Bennett's acceptance ratio method (83-85). In this way, we can estimate $W_{rev}$ that is the work that we would measure if we were able to perform the experiment under reversible conditions. In such case, averaging over infinite realizations, we would observe the TFEC (58), which we now set out to reconstruct.

First, we need to characterize the elastic response of the DNA-handle/protein system. This is simply achieved by fitting the FECs with two quadratic polynomials as showed in Figure 3. Note that the elastic response of the DNA handles is usually modeled with an extensible worm-like chain. However, for the purpose of the TFEC reconstruction, given the short span of forces in our integration region, a humble parabolic fit is the simplest tool that will do the job. Once we have a good enough representation of the native and denatured branches, we turn our attention to the rip. We know that the slope along the rip is equal to minus the stiffness of the trap, which we can measure as the ratio between the loading rate and the pulling speed. Then the only missing detail is where exactly the rip should be placed. It turns out that the request that the area under the TFEC matches the value $W_{rev}$, which we already know because of Crooks' theorem, unequivocally determines the position of the rip, and thus completes the reconstruction of the TFEC. The uncertainty introduced by such procedure has been estimated by means of a statistical bootstrap: knowing the variance associated to Bennett's estimator of $W_{rev}$ (85), we repeat 100 times the TFEC reconstruction starting from 100 values of the reversible work sampled from a Gaussian distribution. This is the origin of the errors in Tables 2 and 3.

This procedure for building the TFEC from experimental data has, to the best of our knowledge, never been proposed before. As it might be useful to other researchers who employ a similar setup, we plan to thoroughly discuss its validation in a forthcoming paper.

We can now turn our attention to the region of the TFEC that contains the rip. To fix the notation, let's say that the protein breaks at ($x_0$, $f_0$) and that the rip ends at ($x_1$, $f_1$) (see Figure S2 in the Supplementary Information). With these conventions, the coexistence force is $f_c = (f_0+f_1)/2$, and the extension released upon unfolding is $\Delta x_c = x_1 - x_0$. The product $f_c \cdot \Delta x_c = \Delta G$ is the area below the rip and represents the equilibrium work that we need to put into the system in order to 1) denature the protein and 2) stretch it from zero to the equilibrium length expected for the denatured protein at $f = f_1$. To round up the result, we also need to take into account the loss of entropy due to partial orientation of the protein in its native state when $f = f_0$ (86) and the fact that the extension of the handles decreases along the rip to accommodate for the stretching of the denatured protein. When all contributions have been dutifully considered, we are left with an estimate of $\Delta G_0$, the free energy of denaturation of the protein. The route from $\Delta G$ to $\Delta G_0$, however, has been already discussed in detail elsewhere (46), so we recall it briefly in the Supplementary Information.

**Key words:** ACBP, ligand binding, optical tweezers, energy landscape, coenzyme-A, protein unfolding.

**Author Contributions**

P.S. performed experiments and data analysis; L.B. performed MD simulations and wrote the article; A.M. performed thermodynamics force-extension curve reconstruction, data analysis and wrote the article; P.O.H. designed research, synthesized protein-DNA molecular constructs and wrote the article; B.B.K. contributed protein samples and wrote the article; C.C. designed research and wrote the article.

**Acknowledgments**

P.O.H. and B.B.K. acknowledge the Carlsberg Foundation, the Lundbeck Foundation and Novo Nordisk Foundation for financial support. C.C. gratefully acknowledges the University of Modena and Reggio Emilia for financial support through the grant 020145_17_FDA_CARRAFAR2016INTER.

**Figures**

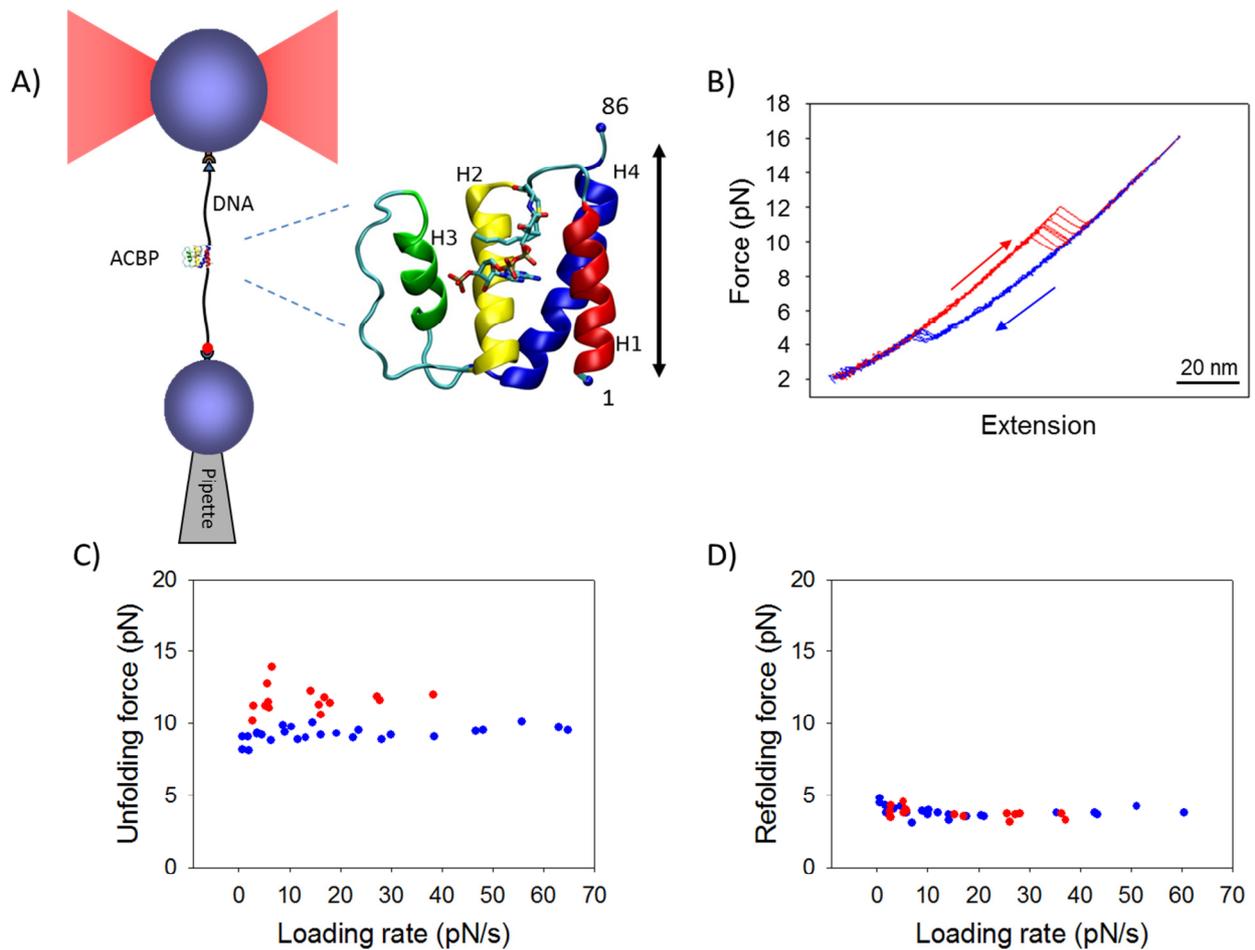

**Figure 1**. Mechanical optical tweezers manipulation of ACBP in the presence of octanoyl-CoA. A) Experimental setup. DNA handles (500 bp) covalently attached to two cysteine residues are used to specifically tether the protein to two polystyrene beads via biotin/streptavidin or digoxigenin/antibodies interactions (13, 45). The force applied on the molecule is varied by moving the pipette relative to the optical trap. NMR structure of ACBP (Protein Data Bank code 1NTI) bound to octanoyl-CoA is shown in the inset (49, 87). B) Overlaid force vs extension cycles obtained by stretching and relaxing the protein multiple times. Each stretching trace (red) shows a discontinuity (rip) around 10-12 pN due to the sudden increase in extension of the protein upon unfolding, as it passes from a compact native state to an elongated stretched unfolded state. Each relaxation trace shows a rip in the opposite direction due to compaction of the protein upon refolding. Fitting the worm-like chain (WLC) model (19, 88) to the stretching traces yielded a $\Delta Lc$ of 27 ± 2 nm, which compares well with the theoretical $\Delta Lc$ of 28.2 nm calculated as described previously (45). C) Most likely unfolding forces measured at different loading rates while pulling

on ACBP in the presence (red dots) or in the absence (blue dots) of octanoyl-CoA (89, 90). D) Most likely refolding forces measured at different loading rates while pulling on ACBP in the presence (red dots) or in the absence (blue dots) of octanoyl-CoA (89, 90).

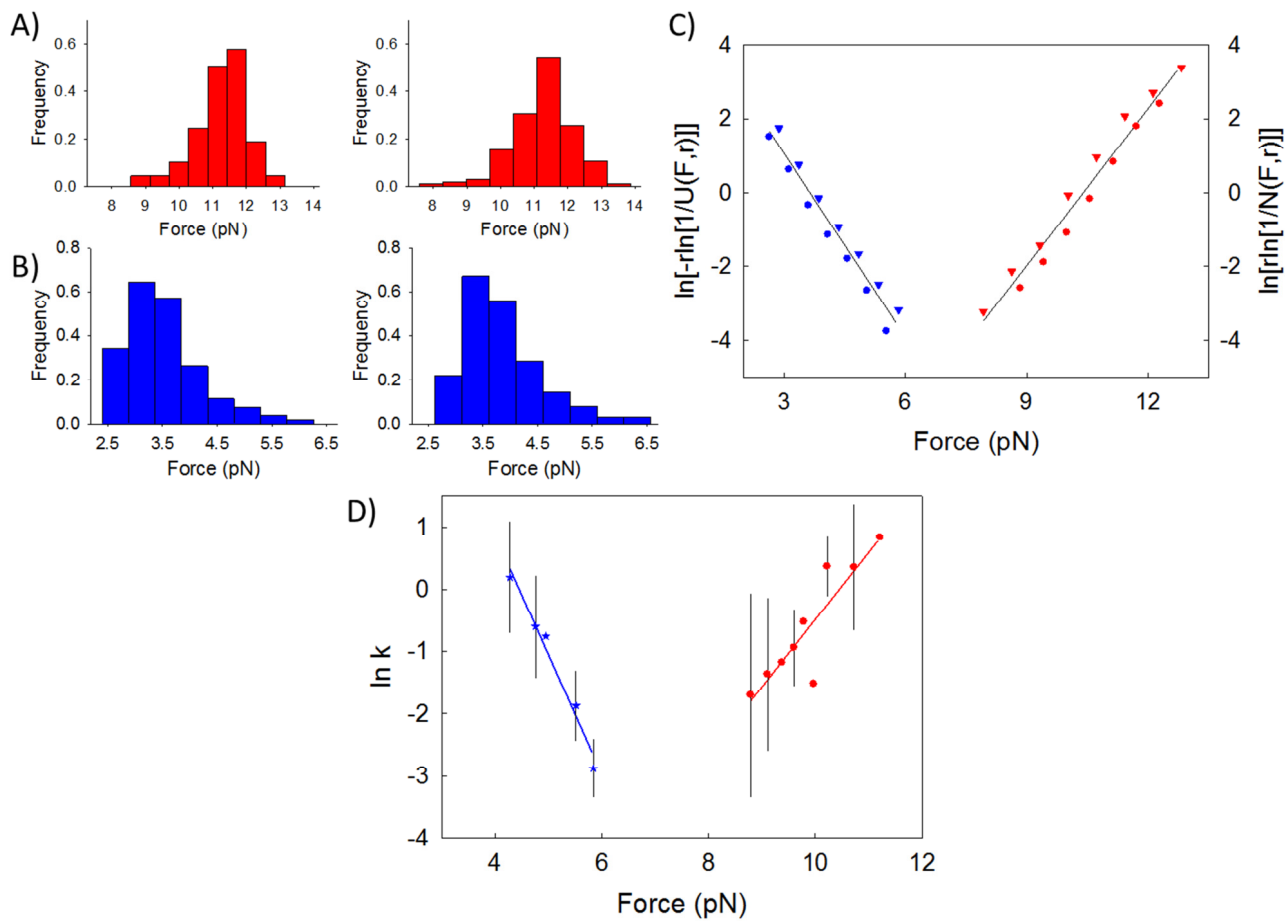

**Figure 2**. Kinetics of ACBP folding in the presence of octanoyl-CoA. A) Unfolding force distributions obtained in force-ramp experiments. Left panel: loading rate $r = 6$ pN s$^{-1}$, $N = 150$; right panel: $r = 3$ pN s$^{-1}$, $N = 121$. B) Refolding force distributions obtained in force-ramp experiments. Left panel: loading rate $r = -2$ pN s$^{-1}$, $N = 109$; right panel: $r = -3$ pN s$^{-1}$, $N = 127$. C) Plots of $ln[r\ ln[1/N]]$ and $ln[-r\ ln[1/U]]$ vs force, where $N$ and $U$ are the folded and unfolded fractions, respectively. Red triangles and red circles are from the left and right panels of A), respectively. Blue circles and blue triangles are from the left and right panels of B), respectively. Data acquired at different loading rates overlap. The best fit values for $x^{\ddagger}_u$, $x^{\ddagger}_f$, $k^0_u$ and $k^0_f$ are reported in Table 1. D) Unfolding (red) and refolding (blue) rates measured in force-jump experiments (45, 51). Experimental data were fit to the Bell model to estimate the position of the transition state and the rate constants at zero force (Table 1).

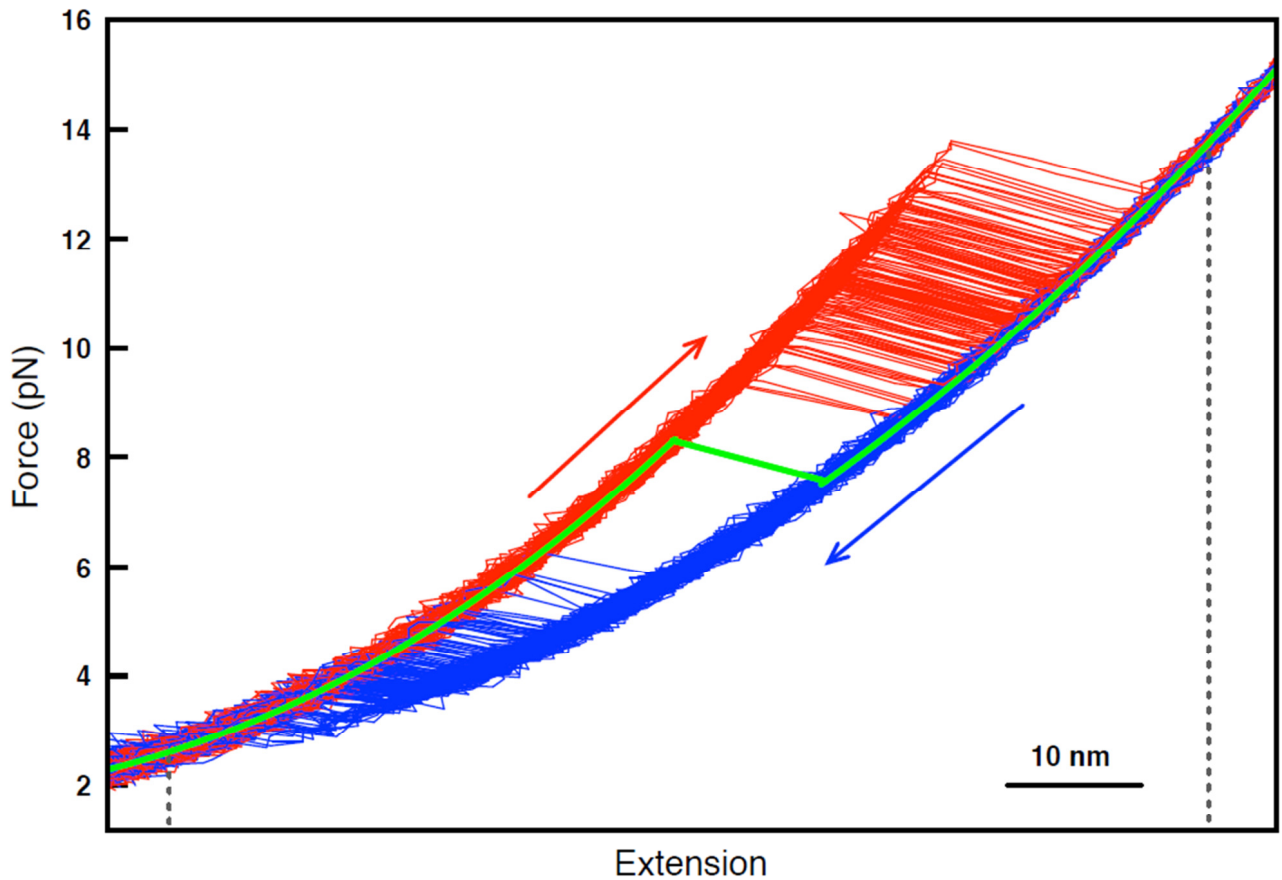

**Figure 3**. Thermodynamics of ACBP folding in the presence of octanoyl-CoA. Here, all the 158 cycles of unfolding/refolding FECs belonging to the same dataset are represented together with the Thermodynamic Force-Extension Curve (TFEC, solid light green line), which is the FEC we would obtain if we were to average over infinite realizations of the experiment performed under reversible (i.e., quasi-static) conditions. The dashed grey vertical lines indicate the integration region selected to compute the work. The slope along the rip is the negative of the trap stiffness, while the position of the rip is unequivocally determined by the request that the area under the TFEC equals the reversible work estimated by means of the Crooks fluctuation theorem (55).

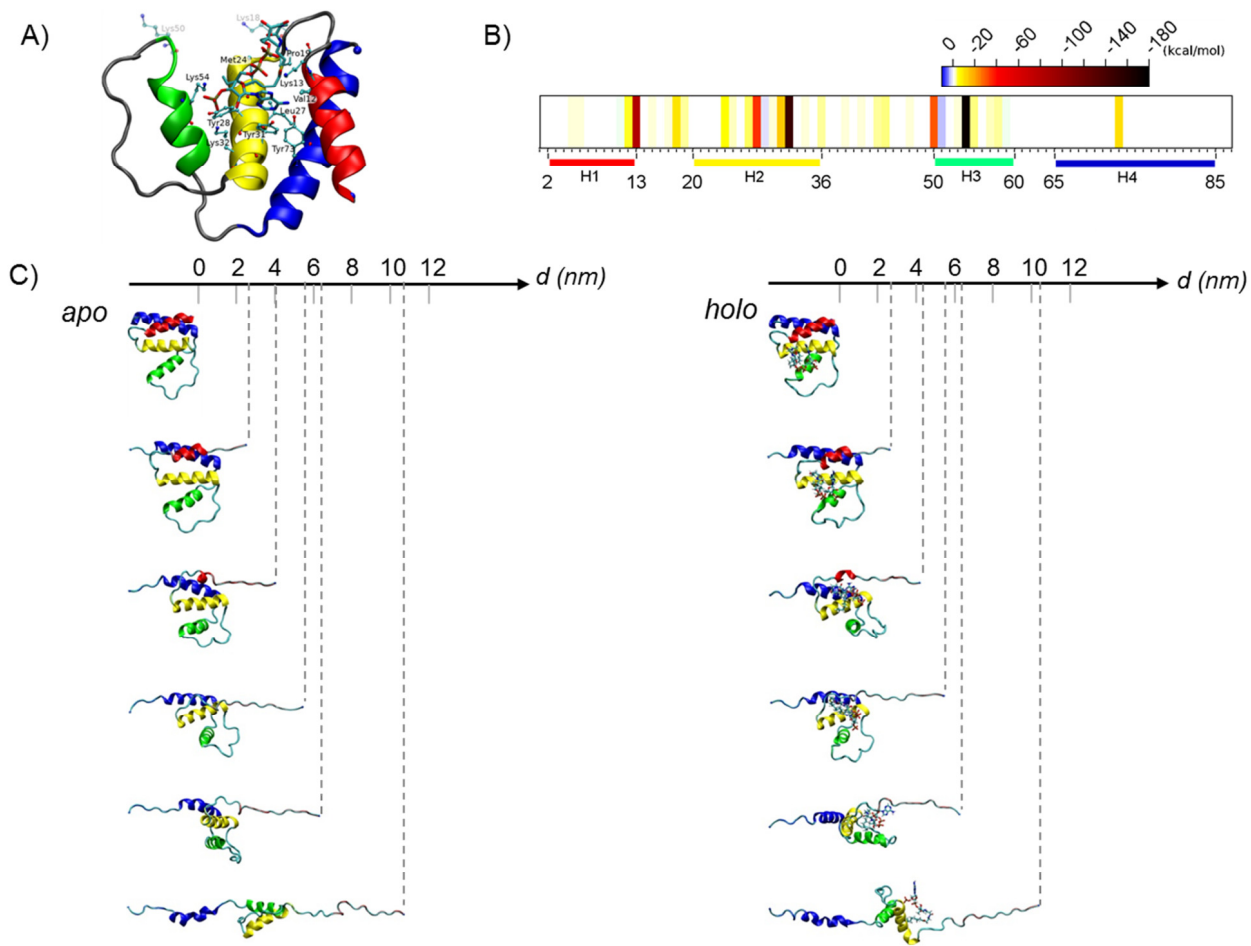

**Figure 4**. Equilibrium and SMD simulations of ACBP unfolding. A) Representative MD simulation structure of holo ACBP, showing the residues mainly involved in the interaction with the ligand octanoyl-CoA. B) Average interaction potential energies between residues of ACBP and octanoyl-CoA. Weak unfavorable interactions (positive values) are shown in light blue and were limited to the Ser29 and Gly51, which do not interact directly with the ligand. Weak dispersion and hydrophobic interactions are indicated in yellow (from 0 to -10 kcal/mol), while hydrogen bonds, polar and electrostatic interactions are shown in red (from -10 to -100 kcal/mol). Dark colors indicate strong ionic/electrostatic interactions. C) Snapshots of the simulated unfolding trajectories of the apo and holo forms of ACBP selected at different molecular extensions. Helix color scheme is the same as in A).

| Table 1. Position of *TS* and reaction rates at zero force | | | | | | | |
|---|---|---|---|---|---|---|---|
| **Protein** | **Experiment** | $x^{\ddagger}_u$ (nm) | $x^{\ddagger}_f$ (nm) | $k_m k^0_u$ (s$^{-1}$) | $k_m k^0_f$ (s$^{-1}$) | $m\beta_{Tu}$ | $m\beta_{Tf}$ |
| Holo ACBP | Force-ramp | 5.8 ± 0.4 | 6.8 ± 0.4 | 6.0(±4)·10$^{-7}$ | 1.0(±0.4)·10$^3$ | 0.45±0.7 | 0.55±0.7 |
|  | Force-jump | 6.0 ± 0.73 | 7.9 ± 0.73 | 2.9(±5.3)·10$^{-7}$ | 5.4(±4.9)·10$^3$ | | |

**Table 2.** Results of the TFEC reconstruction for holo ACBP. Here $\Delta G_0$ is the zero-force free energy difference between the fully denatured apo ACBP and the native holo ACBP states. Each data set was acquired from 1 molecule

|  | $\Delta x_c$ (nm) | $f_c$ (pN) | $\Delta G_0/(k_BT)$ |
|---|---|---|---|
| dataset 1 (118 cycles) | 11.4 ± 0.6 | 7.5 ± 0.5 | 14.3 ± 1.7 |
| dataset 2 (131 cycles) | 11.4 ± 0.8 | 7.7 ± 0.6 | 14.4 ± 2.3 |
| dataset 3 (150 cycles) | 11.1 ± 1.0 | 7.0 ± 0.8 | 13.3 ± 3.5 |
| dataset 4 (159 cycles) | 11.0 ± 0.9 | 7.6 ± 0.8 | 13.8 ± 3.9 |
| dataset 5 (109 cycles) | 10.9 ± 1.2 | 6.5 ± 0.8 | 12.5 ± 4.9 |
| **Weighted Average** | **11.2 ± 0.4** | **7.4 ± 0.3** | **14.1 ± 1.2** |

**Table 3.** Results of the TFEC reconstruction for apo ACBP. Here $\Delta G_0$ is the zero-force free energy difference between the fully denatured apo ACBP and the native apo ACBP states. Each data set was acquired from 1 molecule

|  | $\Delta x_c$ (nm) | $f_c$ (pN) | $\Delta G_0/(k_BT)$ |
|---|---|---|---|
| dataset 1 (199 cycles) | 8.6 ± 0.4 | 5.8 ± 0.3 | 7.7 ± 0.8 |
| dataset 2 (284 cycles) | 8.0 ± 0.3 | 5.7 ± 0.3 | 6.6 ± 0.5 |
| dataset 3 (286 cycles) | 7.9 ± 0.3 | 6.0 ± 0.3 | 6.5 ± 1.4 |
| dataset 4 (100 cycles) | 7.6 ± 0.7 | 5.2 ± 0.6 | 6.5 ± 1.8 |
| **Weighted Average** | **8.1 ± 0.2** | **5.8 ± 0.2** | **6.9 ± 0.4** |